\magnification \magstep1
\raggedbottom
\openup 2\jot
\hsize=15truecm
\vsize=23truecm
\voffset6truemm
\headline={\ifnum\pageno=1\hfill\else
\hfill {\it Singolarit\`a Cosmologiche} \hfill \fi}
\centerline {\bf SINGOLARIT\`A COSMOLOGICHE}
\vskip 1cm
\centerline {Giampiero Esposito}
\vskip 1cm
\noindent
{\it Istituto Nazionale di Fisica Nucleare, Sezione
di Napoli, Mostra d'Oltremare, Padiglione 20, 80125 Napoli}.
\vskip 0.3cm
\noindent
{\it Universit\`a degli Studi di Napoli Federico II,
Dipartimento di Scienze Fisiche, Mostra d'Oltremare,
Padiglione 19, 80125 Napoli}.
\vskip 1cm
\noindent
Le conoscenze umane sul piano scientifico hanno sempre
proceduto, gradualmente, mosse dalla curiosit\`a e 
dall'ansia di comprendere le ragioni ultime delle leggi
fisiche e matematiche che regolano l'evoluzione dei
fenomeni naturali. La meccanica newtoniana condusse, in
particolare, a degli sviluppi di eccezionale portata in
meccanica celeste, e le menti pi\`u brillanti dei secoli
passati elaborarono degli schemi di calcolo e dei teoremi
matematici per analizzare in dettaglio il moto dei corpi
celesti. A partire da una legge di forza la cui formulazione
era semplice, la mente umana si \`e avventurata, in un'epoca
in cui non erano disponibili i moderni strumenti di calcolo,
nello studio del problema dei 3 corpi, delle maree, del moto
della luna, nel calcolo delle orbite dei pianeti. I moderni
sviluppi, come ad esempio il satellite al guinzaglio (e qui
ricordiamo il contributo del compianto Professor Giuseppe
Colombo) e le missioni spaziali, sono la continuazione di
quegli sforzi che hanno tenuto impegnati fisici e matematici
da Galileo sino ai giorni nostri. 

In altri rami della fisica,
fu una grande conquista l'unificazione dei fenomeni elettrici
e magnetici attraverso l'introduzione del concetto di campo
elettromagnetico. Le corrispondenti equazioni, che portano il
nome di James Clerk Maxwell, descrivono una vasta messe di
dati e fenomeni, e hanno ricevuto ampia conferma dagli 
esperimenti. D'altronde, fu proprio da un profondo riesame
critico delle leggi della meccanica, dell'elettromagnetismo,
e dei principi di invarianza, che prese le mosse, all'inizio 
di questo secolo, la teoria della relativit\`a di Albert
Einstein. I contributi di Poincar\'e, Einstein, 
Ricci-Curbastro, Levi-Civita e Marcel Grossmann portarono ad
una profonda revisione del modo in cui i fenomeni gravitazionali
venivano compresi e descritti. La gravitazione \`e una delle 4
interazioni fondamentali oggi note. La sua caratteristica
peculiare \`e di essere sempre attrattiva, ed \`e la principale
responsabile della struttura su larga scala dell'universo in
cui viviamo. Grazie agli studiosi prima menzionati, divenne
chiaro che tutti i fenomeni gravitazionali sono intimamente
correlati a propriet\`a geometriche, le quali traducono in maniera
chiara e semplice le richieste di invarianza che animano le
moderne teorie fisiche. I fenomeni gravitazionali sono
regolati da equazioni matematiche le quali, pur se tecnicamente
complicate, possono essere risolte in casi fisicamente 
rilevanti. Qui interviene, per\`o, un fatto nuovo. Prima della
teoria di Einstein della gravitazione, la situazione familiare
in fisica era che le equazioni matematiche del problema, assieme 
ad opportune condizioni supplementari (e.g. il comportamento
all'infinito, o agli estremi di un intervallo, o su superfici di
contorno) determinava completamente e univocamente l'oggetto
di indagine del fisico teorico. In altri termini, le leggi della
fisica erano sempre valide, ed esse bastavano a determinare
l'evoluzione di tutti i fenomeni osservabili. Tuttavia, a met\`a
degli anni sessanta (circa mezzo secolo dopo che Einstein ebbe
completato la sua teoria della ``relativit\`a generale") 
cominci\`o a diventare chiaro che la teoria di Einstein prevede,
in modo alquanto generico, che una ``singolarit\`a cosmologica"
deve essere esistita. Questo \`e un concetto cruciale al quale
\`e dedicato il resto della nostra relazione. 

Anzitutto, una mente curiosa e educata a esercitare il suo
spirito critico vorrebbe comprendere ``cosa sia" una 
singolarit\`a cosmologica, prima ancora di lanciarsi in ardite
analisi matematiche per dimostrare la sua assenza o la sua
presenza in cosmologia. A onor del vero, il concetto era di
cos\`\i\ difficile comprensione che anche gli studiosi che
diedero i maggiori contributi a questo campo di ricerche (in
ordine cronologico: Penrose, poi Hawking, poi Geroch) ebbero 
non poche difficolt\`a a formulare in maniera precisa una
definizione che catturasse tutte le patologie di interesse
fisico. L'esperienza con problemi pi\`u semplici suggerisce
di riguardare come ``singolare" una situazione in cui una
o pi\`u grandezze di interesse fisico diventano infinite. 
Per le interazioni gravitazionali, si potrebbe allora pensare
a punti di densit\`a infinita, che, se presenti nel problema
dell'origine dell'universo, corrisponderebbero ad una origine
del tempo, a partire da uno stato di densit\`a infinita nel
quale non sarebbe stato possibile imporre la validit\`a di
alcuna delle leggi fisiche a noi note. I risultati matematici
di Penrose, Hawking e Geroch considerarono invero un approccio
pi\`u elegante e profondo. L'idea alla base della definizione 
di singolarit\`a da loro usata era la seguente. Nella teoria 
di Einstein della gravitazione, gli osservatori in moto libero
si muovono lungo una particolare famiglia di curve, le
{\it geodetiche di tipo tempo}. Inoltre, le ``storie" di
particelle con massa a riposo nulla sono rappresentate da
un'altra classe di curve, le {\it geodetiche di tipo luce}.
L'espressione matematica di tali curve involve un parametro,
il {\it parametro affine}, e la possibilit\`a di estendere
tali curve a valori arbitrari di tale parametro corrisponde
al caso in cui gli osservatori in moto libero o le particelle
esistono per sempre. Diversamente, si dovrebbe ammettere che
tali osservatori o particelle esistono solo per un intervallo
temporale finito, per poi sparire misteriosamente alla nostra
vista. Questo evento ``drammatico" appare ancora pi\`u patologico
del caso in cui l'intensit\`a della interazione gravitazionale
dovesse diventare infinita.

Pertanto, si propose di considerare {\it condizioni minimali}
per la assenza di singolarit\`a (cosmologiche) le propriet\`a
di completezza geodetica di tipo tempo e luce. In altri termini,
la teoria di Einstein unifica spazio e tempo in un continuo
a 4 dimensioni. Questo ente unificato prende il nome di
{\it variet\`a spazio-tempo}. Per definizione, lo
spazio-tempo, che \`e l'oggetto ove si svolgono tutti gli
eventi fisici, esibisce completezza geodetica di tipo tempo e
luce se tutte le curve geodetiche di tipo tempo e luce possono
essere estese a valori arbitrari del loro parametro affine.
Un tale modello di spazio-tempo viene detto {\it libero da
singolarit\`a}. Se tale propriet\`a non vale, si dice invece
che lo spazio-tempo \`e singolare, e dunque in esso devono
verificarsi eventi patologici altamente non banali, quali ad
esempio il caso di osservatori in moto libero la cui ``storia"
non esiste dopo, o prima, di un intervallo finito di tempo
(il ``tempo proprio"). La formulazione matematica precisa di
tale idea esula dagli scopi della nostra relazione, ma
dobbiamo avvertire il lettore che questa idea, pur se provatasi
utile nei tardi anni sessanta nel dimostrare i teoremi sulle
singolarit\`a, era essa stessa criticabile e migliorabile. 
Al di l\`a delle generalizzazioni e 
dei raffinamenti tecnici, restava
poco chiaro se le singolarit\`a predette dai teoremi di Hawking
e Penrose fossero riconducibili unicamente a casi in cui
l'intensit\`a del campo gravitazionale diventa infinita.

Cerchiamo, intanto, di cominciare a mettere ordine fra le tante
idee e problemi che stiamo presentando. Le condizioni sotto le
quali i teoremi di Hawking e Penrose sulle singolarit\`a vengono
dimostrati sono essenzialmente le seguenti:
\vskip 0.3cm
\noindent
(i) La gravitazione \`e attrattiva (questa, ricordiamo, \`e una
propriet\`a sempre valida delle interazioni gravitazionali);
\vskip 0.3cm
\noindent
(ii) C'\`e abbastanza materia in qualche regione da impedire a
qualunque cosa di scappare da quella regione;
\vskip 0.3cm
\noindent
(iii) Non vi dovrebbero essere violazioni di causalit\`a (ad
esempio, viaggi indietro nel tempo, sino a influenzare la
vita dei nostri genitori).
\vskip 0.3cm
\noindent
\`E un fatto degno di nota che la teoria di Einstein, di per se
stessa, {\it non} nega a priori la possibilit\`a che qualche
violazione di causalit\`a, che sembrerebbe strana per il
``buon senso" comune, possa invece verificarsi. Nella formulazione
di teorie fisiche, la richiesta (iii) \`e dunque un ingrediente
che va eventualmente {\it aggiunto} dal fisico teorico. Tuttavia,
un teorema sulle singolarit\`a, dovuto a Hawking, vale 
indipendentemente da qualunque ipotesi fatta sulla causalit\`a
o sue possibili violazioni. 

La scoperta teorica che le singolarit\`a cosmologiche devono
presentarsi in modo generico, {\it se} vale la teoria di 
Einstein della gravitazione, fu particolarmente sconvolgente,
in quanto si era a lungo pensato che tali singolarit\`a 
fossero legate all'alto grado di simmetria che alcune soluzioni
particolari delle equazioni di Einstein (e.g. i modelli di
Friedmann, Lemaitre, Robertson e Walker) presentano. In questa
direzione si erano mossi gli studiosi sovietici Lifshitz e
Khalatnikov, ma successivamente questi stessi autori, in
collaborazione con Belinsky, fecero un'analisi pi\`u 
approfondita, che avvalorava le eleganti analisi astratte di
Penrose, Hawking e Geroch: le pi\`u generiche soluzioni 
conducono ancora a modelli spazio-temporali singolari. Sembra
dunque chiaro, da quanto sin qui esposto, che le concezioni
cosmologiche, e pi\`u in generale tutto l'apparato della
fisica teorica faticosamente costruito da generazioni di
studiosi, venivano a trovarsi di fronte ad una crisi senza
precedenti. Infatti i teoremi prima ricordati, e i loro
raffinamenti successivi, indicavano che, nell'ambito della
teoria di Einstein dei fenomeni gravitazionali, le 
singolarit\`a, riguardate come punti ove le equazioni di 
Einstein e le altre leggi fisiche note non sono pi\`u valide,
sono propriet\`a del tutto generiche. In altri termini, la
fisica ci presenterebbe, da un lato, un insieme di leggi
matematiche chiare e ben definite, che prevedono con successo
una gran messe di dati empirici, mentre poi ci indicherebbe
che, in certi regimi estremi, questo insieme di regole e
principi non \`e pi\`u valido, n\'e \`e chiaro come 
sostituirlo con altre regole e altri principi.

Questa osservazione prepara il terreno per l'ultima parte del\-la
nostra re\-la\-zio\-ne, che pren\-de le mos\-se da un 
la\-vo\-ro sci\-en\-ti\-fi\-co
pre\-sen\-ta\-to da Hawking all'at\-ten\-zio\-ne del\-la Pontificia Accademia
delle Scienze. Si era allora nel 1982, e lo studioso britannico,
che reggeva e regge ancor oggi la celebre cattedra lucasiana
che fu di Newton, era pi\`u che mai ossessionato dal problema
delle singolarit\`a e del loro ruolo in un possibile quadro
della ``teoria finale di ogni cosa" a cui aspira, forse con
troppa ambizione, la fisica teorica dei nostri tempi. 
L'intervento di Hawking era centrato sul problema delle
``condizioni al contorno per l'universo", e su dove esse
andrebbero imposte. Invero, scopo ultimo della fisica \`e di
fornire un modello matematico dell'universo che concordi con le
osservazioni fatte sinora e, inoltre, 
pred\`\i ca i risultati di
future osservazioni. Risulta allora conveniente dividere in 
due parti le formulazioni degli attuali modelli teorici:
\vskip 0.3cm
\noindent
(i) Un insieme di equazioni matematiche che governano le
variabili della teoria;
\vskip 0.3cm
\noindent
(ii) Un insieme di ``condizioni al contorno" per le soluzioni
di tali equazioni. Tali condizioni specificano, come anticipato
all'inizio della nostra relazione, il comportamento delle 
variabili della teoria sulle superfici di contorno che si
presentano in natura.
\vskip 0.3cm
\noindent
Questa separazione della fisica in equazioni matematiche con
relative condizioni al contorno pu\`o apparire artificiosa, ma
si \`e invero rivelata utile nel condurre ad un progresso delle
conoscenze. Infatti, la situazione tipica \`e quella in cui le
osservazioni ci consentono una conoscenza solo {\it locale}
dell'universo. \`E molto pi\`u difficile costruire modelli
teorici che rendano conto dell'universo nella sua interezza e
senza alcun bisogno della specificazione di condizioni
supplementari. Tuttavia, per avere una teoria soddisfacente
dell'universo (ove mai possibile), resta allora il problema
di specificare le condizioni al contorno in modo appropriato.

Man mano che ci si avvicina alla scala di energie caratteristica
della possibile singolarit\`a iniziale dell'universo, si devono
considerare quegli ingredienti aggiuntivi delle teorie fisiche
che rientrano nell'ambito della fisica quantistica. Qui interviene
un ulteriore, profondo concetto. Il ``collasso" della fisica 
implicato dalla singolarit\`a cosmologica della teoria di
Einstein viene oggi interpretato come una indicazione che non
\`e la fisica a perdere di validit\`a. Appare invece pi\`u
plausibile che sia stata applicata la teoria di Einstein al di
l\`a del suo limite di validit\`a naturale. Alle altissime
energie in gioco nell'universo primordiale non si possono non
considerare gli effetti della ``meccanica quantistica". Non
esiste ancora accordo tra le varie ``scuole" della fisica
teorica contemporanea su come combinare in una sintesi superiore
la visione di Einstein della gravitazione con i principi della
meccanica quantistica. Tuttavia, la ricerca \`e ormai orientata
verso un superamento della teoria di Einstein in senso stretto,
cercando di formulare una teoria pi\`u profonda che si riduca a
quella di Einstein quando le energie in gioco sono quelle della
nostra esperienza quotidiana. Nel suo contributo del 1982, e
negli anni a seguire, Hawking osserv\`o che, a causa delle
possibili ``fluttuazioni" nella struttura dello spazio-tempo
alle altissime energie, non \`e sufficiente imporre condizioni
al contorno in corrispondenza della singolarit\`a iniziale,
anche se tali condizioni dovessero 
diventare ben definite grazie alla
fisica quantistica. Egli {\it propose} invece di descrivere
l'universo in termini di un continuo a 4 dimensioni 
{\it senza alcuna superficie di contorno}. A questo modello
matematico, che non \`e imposto, si badi bene, dalle leggi
fisiche oggi note, corrisponderebbe un universo fisico che 
non ha origine n\'e fine, privo di singolarit\`a iniziale.
In tale universo, le leggi fisiche determinerebbero la
totalit\`a degli eventi osservabili e predicibili, e non si
avrebbe alcun collasso della fisica. 

Questo scenario teorico, che, non ci stancheremo di ripeterlo,
rappresenta solo uno dei possibili schemi matematici 
elaborati dalla mente umana, unito alle idee e ai risultati 
discussi in precedenza, solleva interrogativi profondi e
irrisolti:
\vskip 0.3cm
\noindent
(i) Se, a livello classico, le singolarit\`a cosmologiche
sono una propriet\`a generica della teoria di Einstein della
gravitazione, si pu\`o concludere che, a livello quantistico, 
le singolarit\`a vengono eliminate in modo generico? O 
invece permangono anche a livello quantistico, salvo a 
seguire approcci lungo le linee di Hawking e della sua
scuola?
\vskip 0.3cm
\noindent
(ii) Cosa ci insegnano le propriet\`a e i modelli 
matematici discussi sinora
su quel che \`e alla base del cosmo e delle sue leggi?
\vskip 0.3cm
\noindent
(iii) L'universo di Hawking, senza superfici di bordo,  
traduce davvero l'idea di un cosmo
senza origine n\'e fine, in cui il tempo non ha mai avuto
inizio?
\vskip 0.3cm
\noindent
(iv) Qual \`e il significato ultimo (se pure esiste) dei 
nostri sforzi per una migliore comprensione dei fenomeni
naturali? Lo scopo ultimo dell'esistenza e della ricerca
umana \`e forse il completamento di un processo di
auto-conoscenza, in cui sviluppiamo gradualmente (lungo un
arco di secoli, forse di millenni) la sensibilit\`a per
quelle strutture matematiche mediante le quali 
reinterpretare in modo limpido e chiaro le leggi che regolano
l'universo e la sua evoluzione?
\vskip 0.3cm
\noindent
Questi interrogativi sembrano confermare che non esiste pi\`u
una netta separazione tra il ruolo dello scienziato e quello
del filosofo, e che entrambi, pur se mossi da punti di vista
molto differenti, possono trovare un terreno comune. 
Mentre le nostre conoscenze progrediscono,
mentre nuove equazioni ci schiudono le porte di decenni di
nuove ricerche affascinanti, cresce in tutti noi la
curiosit\`a e l'ansia di trovare risposte a tali domande. Ed
\`e con questa curiosit\`a infinita che le generazioni di
studiosi presenti e future possono legittimamente aspirare
ad arricchire la loro esistenza, nella speranza che questi 
sforzi incessanti ci aiutino a comprendere l'origine e il
fine ultimo dell'universo e delle sue leggi.
\vskip 1cm

\parindent=0pt
\everypar{\hangindent=20pt \hangafter=1}
\leftline {\bf Bibliografia}
\vskip 1cm

[1] Beem J. K. , Ehrlich P. E. e Easley K. L. (1996)
{\it Global Lorentzian Geometry} (New York: Dekker).

[2] Belinsky V. A. , Khalatnikov I. M. e Lifshitz E. M. (1970)
Oscillatory Approach to a Singular Point in Relativistic
Cosmology, {\it Adv. in Phys.} {\bf 19}, 523--573.

[3] Clarke C. J. S. (1975) Singularities in Globally 
Hyperbolic Space-Times, {\it Commun. Math. Phys.}
{\bf 41}, 65--78.

[4] Clarke C. J. S. (1976) Space-Time Singularities,
{\it Commun. Math. Phys.} {\bf 49}, 17--23.

[5] Clarke C. J. S. e Schmidt B. G. (1977) Singularities:
The State of the Art, {\it Gen. Rel. Grav.} 
{\bf 8}, 129--137.

[6] Clarke C. J. S. e Krolak A. (1985) Conditions for the
Occurrence of Strong Curvature Singularities, {\it J. Geom.
Phys.} {\bf 24}, 127--143.

[7] Clarke C. J. S. (1993) {\it The Analysis of Space-Time Singularities},
Cambridge Lecture Notes in Physics, Vol. 1 (Cambridge: Cambridge
University Press).

[8] Esposito G. (1994) {\it Quantum Gravity, Quantum Cosmology and
Lorentzian Geometries}, Springer Lecture Notes in Physics, New
Series m: Monographs, Vol. m12 (Berlin: Springer-Verlag).

[9] Geroch R. P. (1966) Singularities in Closed Universes, 
{\it Phys. Rev. Lett.} {\bf 17}, 445--447.

[10] Geroch R. P. (1968a) Local Characterization of Singularities
in General Relativity, {\it J. Math. Phys.} {\bf 9}, 450--465.

[11] Geroch R. P. (1968b) What is a Singularity in General Relativity? , 
{\it Ann. Phys.} {\bf 48}, 526--540.

[12] Geroch R. P. (1970) Singularities, in {\it Relativity},
eds. S. Fickler, M. Carmeli e L. Witten (New York: Plenum Press)
259--291.

[13] Halliwell J. J. (1991) Quantum Cosmology and the Creation
of the Universe, {\it Scientific American}, 
December issue, 28--35.

[14] Hawking S. W. (1966a) Singularities and the Geometry of
Space-Time, {\it Adams Prize Essay} (St. John's College, Cambridge).

[15] Hawking S. W. (1966b) The Occurrence of Singularities in
Cosmology, {\it Proc. Roy. Soc. London} {\bf A 294}, 511--521.

[16] Hawking S. W. (1966c) The Occurrence of Singularities in 
Cosmology. II {\it Proc. Roy. Soc. London} {\bf A 295}, 490--493.

[17] Hawking S. W. (1967) The Occurrence of Singularities in
Cosmology. III. Causality and Singularities, 
{\it Proc. Roy. Soc. London} {\bf A 300}, 187--201.

[18] Hawking S. W. e Penrose R. (1970) The Singularities of 
Gravitational Collapse and Cosmology, {\it Proc. Roy. Soc. London}
{\bf A 314}, 529--548.

[19] Hawking S. W. e Ellis G. F. R. 
(1973) {\it The Large-Scale Structure of
Space-Time} (Cambridge: Cambridge University Press).

[20] Hawking S. W. (1982) The Boundary Conditions of the Universe, 
in {\it Pontificiae Academiae Scientiarum Scripta      
Varia} {\bf 48}, 563--574.

[21] Hawking S. W. (1984) The Quantum State of the Universe, 
{\it Nucl. Phys.} {\bf B 239}, 257--276.

[22] Hawking S. W. e Penrose R. (1995) {\it The Nature of Space
and Time} (Princeton: Princeton University Press).

[23] Horowitz G. T. e Myers R. (1995) The Value of Singularities,
{\it Gen. Rel. Grav.} {\bf 27}, 915--919.

[24] Kriele M. (1990a) A Generalization of the Singularity
Theorem of Hawking and Penrose to Space-Times with
Causality Violations, {\it Proc. Roy. Soc. London}
{\bf A 431}, 451--464.

[25] Kriele M. (1990b) Causality Violations and Singularities,
{\it Gen. Rel. Grav.} {\bf 22}, 619--623.

[26] Lifshitz E. M. e Khalatnikov I. M. (1963) Investigations
in Relativistic Cosmology, {\it Adv. in Phys.} {\bf 12},
185--249.

[27] O'Neill B. (1983) {\it Semi-Riemannian Geometry}
(New York: Academic).

[28] Penrose R. (1965) Gravitational Collapse and Space-Time
Singularities, {\it Phys. Rev. Lett.} {\bf 14}, 57--59.

[29] Penrose R. (1979) Singularities and Time-Asymmetry, in
{\it General Relativity, an Einstein Centenary Survey}, eds.
S. W. Hawking e W. Israel (Cambridge: Cambridge University 
Press) 581--638.

[30] Penrose R. (1983) {\it Techniques of Differential 
Topology in Relativity} (Bristol: Society for Industrial 
and Applied Mathematics).

\bye